\documentclass[12pt]{article}
\usepackage{authblk}
\usepackage{hyperref}
\usepackage[top=25truemm,bottom=28truemm,left=24truemm,right=24truemm]{geometry}
\usepackage{setspace}
\normalsize
\usepackage{amsmath,amssymb,mathrsfs,amsbsy,latexsym,amsfonts,amsthm}
\usepackage{fancyhdr}
\pdfoutput=1 % For pdfLaTeX
\usepackage{graphicx}
\usepackage{color}
\usepackage{comment}
\usepackage{here}
\usepackage{braket}
\usepackage[utf8]{inputenc}
\usepackage{mathtools}

\newcommand{\nn}{\nonumber\\}
\newcommand{\bO}{\mathbf{\Omega}}

\DeclareMathOperator{\tr}{tr}
\usepackage{epsf}

\numberwithin{equation}{section}
\title{\bf %Remarks on 
Rindler Bulk Reconstruction\\
and Subregion Duality in AdS/CFT}

\author[1, 2]{
Sotaro~Sugishita\thanks{\tt sugishita.sotaro.r6(at)f.mail.nagoya-u.ac.jp}
}
\author[3]{
	Seiji~Terashima\thanks{\tt terasima(at)yukawa.kyoto-u.ac.jp}
}
\vspace{5mm}
\affil[1]{\it\normalsize Institute for Advanced Research, 
Nagoya University,
Nagoya, Aichi 464-8601, Japan}
\affil[2]{\it\normalsize Department of Physics, 
Nagoya University,
Nagoya, Aichi 464-8602, Japan}
\affil[3]{\it\normalsize 
Center for Gravitational Physics and Quantum Information,  
\mbox{Yukawa Institute for Theoretical Physics, Kyoto University, Kyoto 606-8502, Japan}  }
\date{}
\begin{document}

\maketitle
\thispagestyle{fancy}
\renewcommand{\headrulewidth}{0pt}

%%%%%%%%%%%%%%%%%%%%%%%%%%%%%%%%%%%%%%%%%%%%%%%%%%%%%%%%%%%%%%
%%%%%%%%%%%%%%%%%%%%%%%%%%%%%%%%%%%%%%%%%%%%%%%%%%%%%%%%%%%%%%
\begin{abstract}

In this paper, 
we study the AdS-Rindler reconstruction. 
The CFT operators naively given by 
the holographic dictionary for the AdS-Rindler reconstruction
contain tachyonic modes, which are inconsistent with the causality and unitarity of the CFT.
Therefore, the subregion duality and the entanglement wedge reconstruction do not hold.
We also find that the tachyonic modes in the AdS-Rindler patch lead to arbitrary high-energy or trans-Planckian modes in the global AdS.
It means that the mode expansion of the Rindler patch is sensitive to the UV limit of the theory, that is, quantum gravity.  
In addition, the tachyonic modes are related to the existence of null geodesics connecting the past and future horizons.

\end{abstract}

\newpage
\thispagestyle{empty}
\setcounter{tocdepth}{2}

\setlength{\abovedisplayskip}{12pt}
\setlength{\belowdisplayskip}{12pt}

\tableofcontents
\newpage
%%%%%%%%%%%%%%%%%%%%%%%%%%%%%%%%%%%%%%%%%%%%%%%%%%%%%%%%%%%%%%
%%%%%%%%%%%%%%%%%%%%%%%%%%%%%%%%%%%%%%%%%%%%%%%%%%%%%%%%%%%%%%
\section{Introduction and summary}
%%%%%%%%%%%%%%%%%%%%%%%%%%%%%%%%%%%%%%%%%%%%%%%%%%%%%%%%%%%%%%
%%%%%%%%%%%%%%%%%%%%%%%%%%%%%%%%%%%%%%%%%%%%%%%%%%%%%%%%%%%%%%

It is important to study how the bulk gravitational theory emerges from the CFT in 
the AdS/CFT correspondence
in order to understand what is the spacetime in the quantum gravity.
An explicit realization of this for the bulk fields is called the bulk reconstruction
and has been studied, for example, in \cite{Bena:1999jv, Duetsch:2002hc, Hamilton:2006az, Papadodimas:2012aq, Terashima:2017gmc}, in particular for the free bulk theory limit.

A basic question of the bulk reconstruction is what is the reconstructable
bulk fields from
CFT operators supported only in a subregion of the boundary spacetime.
The subregion duality \cite{Wall:2012uf,Bousso:2012sj, Czech:2012bh, Headrick:2014cta, Almheiri:2014lwa, Jafferis:2015del, Dong:2016eik} 
claims that the bulk fields supported in a subregion of the bulk spacetime, called the entanglement wedge, can be reconstructed from CFT operators on the boundary subregion, but
the bulk fields outside it cannot be reconstructed.
Here the boundary subregion for the CFT operators corresponds to  a boundary limit of the bulk subregion.
This reconstruction is called the entanglement wedge reconstruction and assumed to be correct in many studies
although it is claimed to be incorrect in \cite{Terashima:2020uqu, Terashima:2021klf}.
In particular, for the Rindler patch of the AdS spacetime, the explicit bulk reconstruction formula was given in  
\cite{Hamilton:2006az}
for the free bulk theory limit.
In this AdS-Rindler reconstruction, the boundary limit of the free scalar field on bulk AdS-Rindler is naively identified 
to the CFT primary operator by the BDHM formula \cite{Banks:1998dd}.

In this paper, 
we study the AdS-Rindler reconstruction and find that 
the naive identification %of the boundary limit of the free scalar field on bulk AdS-Rindler to the CFT primary operator 
by the BDHM formula is inconsistent.
Indeed, the CFT operators naively given by 
the BDHM formula for the AdS-Rindler reconstruction
contain tachyonic modes, which are inconsistent with the causality and unitarity of the CFT although these modes are consistent as the bulk theory.\footnote{It is also argued in \cite{Rey:2014dpa} that a difficulty in the bulk reconstruction arising from tachyonic modes in black hole backgrounds (where the modes are called evanescent modes).
In \cite{Leichenauer:2013kaa}, it is also discussed that such tachyonic modes are related to the ill-definedness of the smearing functions in the bulk reconstruction.
}
Here, the important ingredient of this conclusion is that we consider 
the large, but finite $N$ CFT.
Thus, the Planck length (over the AdS-scale) is arbitrary small, but finite.
This means that this inconsistency comes from the truly non-perturbative effects of the quantum gravity.
The free bulk theory, which corresponds to the generalized free CFT, should be
modified above the Planck energy because such a state becomes a black hole and the free spectrum around the fixed background is no longer valid.

In the bulk point of view, there seem to be no problems to consider 
the mode expansion in the AdS-Rindler patch. 
However,
we show that the tachyonic modes in the AdS-Rindler patch correspond 
to arbitrary high energy modes of the global AdS, for example, the trans-Planckian modes.
This means that the mode expansion of the Rindler patch is sensitive to the UV completion of the theory which is the quantum gravity in our case.  
In other words, the low energy modes of the AdS-Rindler patch do not correspond to
the low energy modes of the global AdS. 

Therefore the subregion duality does not hold and the AdS-Rindler reconstruction is incomplete.
It is an important question which part of bulk local fields cannot be reconstructed from the CFT operators in the Rindler patch.
In the AdS-Rindler patch there are null geodesics never reaching the asymptotic boundary. 
This type of null geodesics starts from the past AdS-Rindler horizon and ends on the future one. 
We show that the non-reconstructable tachyonic modes are related to these horizon-horizon geodesics.

Instead of using the AdS-Rindler coordinates,
we can study which part of the bulk local operators are able to be reconstructed by CFT operators in 
a subregion from the global AdS (and the corresponding CFT on the cylinder) viewpoint.
Indeed, in \cite{Terashima:2020uqu, Terashima:2021klf} 
using the 
bulk reconstruction developed in 
\cite{Terashima:2017gmc,Terashima:2019wed, Nagano:2021tbu},
such studies had been done. 
The results obtained in this paper are perfectly consistent with the studies in \cite{Terashima:2020uqu, Terashima:2021klf}.

We believe that the results in this paper are substantial ingredients for the understanding of spacetime in the AdS/CFT and the quantum gravity.
We emphasize that 
the low energy description of the bulk theory with the AdS-Rindler quantization 
should be modified in the AdS/CFT. 
This is interesting because it is often believed that 
the low energy description is valid even in the Rindler coordinate with the horizon because there is no curvature singularity.  
We expect that such a violation is an essential property of (black hole) horizon because it is due to the behavior of fields near the horizon, which is universal to general black hole horizons not restricted to the Rindler one.
This violation might be related to 
the brick wall \cite{tHooft:1984kcu, Iizuka:2013kma}, the fuzzball \cite{Mathur:2005zp, Mathur:2009hf} or the firewall \cite{Almheiri:2012rt} proposals for
the black hole horizon where
the equivalence principle is supposed to be violated although there is no curvature singularity.

We will set the AdS radius $\ell_\text{AdS}=1$ throughout the paper.

%%%%%%%%%%%%%%%%%%%%%%%%%%%%%%%%%%%%%%%%%%%%%%%%%%%%%%%%%%%%%%
%%%%%%%%%%%%%%%%%%%%%%%%%%%%%%%%%%%%%%%%%%%%%%%%%%%%%%%%%%%%%%
\section{Review of AdS-Rindler}\label{sec:Rindler}

In this section, we will review the Rindler patch in the AdS spacetime and the free scalar fields on it.
Some references on the Rindler patch in the AdS/CFT are \cite{Parikh:2012kg, Czech:2012be, Leutheusser:2021qhd, Leutheusser:2021frk}.

%%%%%%%%%%%%%%%%%%%%%%%%%%%%%%%%%%%%%%%%%%%%%%%%%%%%%%%%%%%%%%
%%%%%%%%%%%%%%%%%%%%%%%%%%%%%%%%%%%%%%%%%%%%%%%%%%%%%%%%%%%%%%
\subsection{Coordinates}\label{subsec:cdnt}
We summarize the coordinates of AdS$_{d+1}$ used in this paper.
Using the embedding coordinates into $\mathbf{R}^{2,d}$, AdS$_{d+1}$ is described as 
\begin{align}
    -(X^{-1})^2 -(X^{0})^2 +(X^1)^2+\cdots +(X^d)^2 =-1.
\end{align}

The global coordinates $(\tau,\rho,\Omega)$ are obtained by parameterizing the embedding coordinates as 
\begin{align}
    X^{-1}=\frac{1}{\cos \rho}\cos \tau,\quad   X^{0}=\frac{1}{\cos \rho}\sin \tau,\quad
    X^i=\tan \rho\, \hat{x}^i(\Omega), 
\end{align}
where $\Omega$ represents coordinates of $(d-1)$-dimensional sphere $S^{d-1}$, and $\hat{x}^i(\Omega)$ $(i=1,\dots,d)$ are the embedding of the sphere into $\mathbf{R}^{d}$ as $\sum_{i}(\hat{x}^i)^2=1$.
The coordinates $\tau$ and $\rho$ run in the range $-\infty<\tau<\infty$ and $0\leq \rho<\pi/2$.
In the coordinates, the metric takes
\begin{align}
    ds^2=\frac{1}{\cos^2\! \rho}\left(-d\tau^2+d \rho^2+ \sin^2 \!\rho\, d\Omega_{d-1}^2\right).
\end{align}
For later convenience, we take the  spherical coordinates $\Omega$ as 
\begin{align}
    \hat{x}^1(\Omega)=\cos \theta,\quad
    \hat{x}^j(\Omega)=\sin\theta\,\hat{y}^j(\bO)\quad (j=2, \dots, d) \quad \text{with} \quad 0\leq \theta \leq \pi,
\end{align}
where  $\bO$ represents coordinates of $(d-2)$-dimensional sphere, and $\hat{y}^j(\Omega)$ $(j=2,\dots,d)$ are the embedding of the sphere into $\mathbf{R}^{d-1}$ as $\sum_{j}(\hat{y}^j)^2=1$.\footnote{For $d=2$, we take the range of $\theta$ is $-\pi\leq \theta \leq  \pi$ and $\hat{y}^2(\Omega)=1$.}

\begin{figure}[htbp]
%\vspace{-0.01\columnwidth}
\centering
\includegraphics[width=4cm]{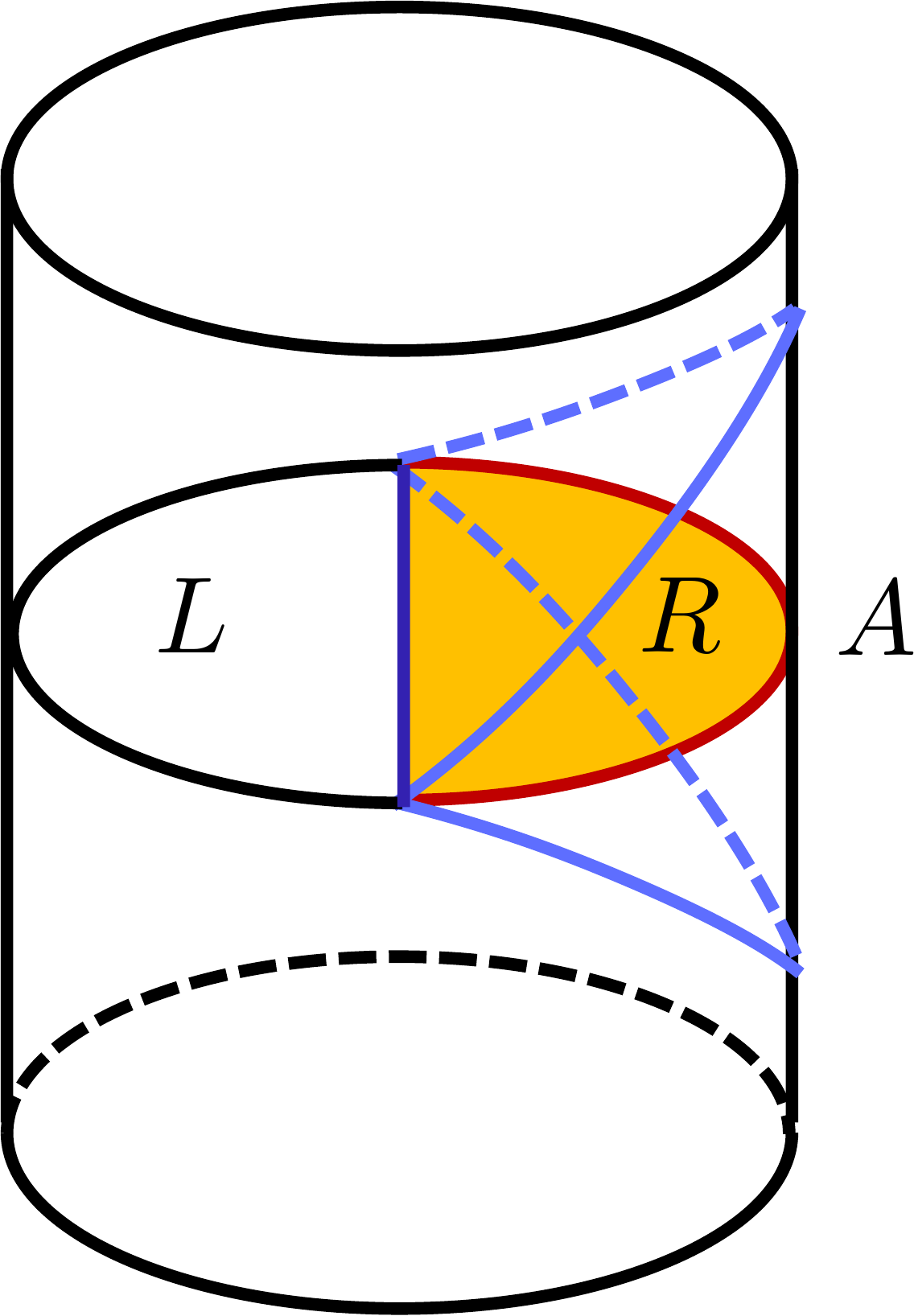}
\caption{AdS-Rindler patch  in the global AdS space.
}
\label{fig:RindAdS}
%\vspace{-0.01\columnwidth}
\end{figure}

We divide AdS$_{d+1}$ as in Fig.~\ref{fig:RindAdS}. A time slice ($\tau=0$) is divided into two subregions $R$ and $L$.
We call the domain of dependence of $R$ (and $L$) the right (left) AdS-Rindler wedge.
%%%%%%%%%%%%%%%%
%%%%%%%%%%%%%%%%
%%%%%%%%%%%%%%%%
%%%%%%%%%%%%%%%%
The coordinates of the right AdS-Rindler wedge $(t_R,\xi,\chi, \bO)$ are given by the parameterization 
\begin{align}
\begin{split}
    &X^{-1}=\sqrt{1+\xi^2}\cosh \chi, \quad X^0=\xi \sinh t_R,\quad  
    X^1=\xi \cosh t_R, \\
    &X^j=\sqrt{1+\xi^2} \sinh \chi\, \hat{y}^j(\bO)\quad (j=2, \dots, d),
\end{split}
\end{align}
with $-\infty<t_R<\infty$, $0\leq\xi<\infty$, $0\leq\chi<\infty$.\footnote{For $d=2$, $-\infty<\chi<\infty$.} 
In these coordinates, the metric becomes
\begin{align}
\label{Rindler-metric}
    ds^2=-\xi^2 dt_R^2+\frac{d\xi^2}{1+\xi^2}+(1+\xi^2) dH_{d-1}^2,
\end{align}
where $dH_{d-1}^2= d\chi^2 +\sinh^2\chi d\Omega_{d-2}^2$ is the metric of $(d-1)$-dimensional hyperbolic space $\mathbf{H}^{d-1}$.
The AdS-Rindler horizon is at $\xi=0$, and the geometry \eqref{Rindler-metric} is called the topological black hole \cite{Emparan:1999gf, Casini:2011kv}.
More precisely, if we introduce $U=X^0-X^1=-\xi e^{-t_R}$ and $V=X^0+X^1=\xi e^{t_R}$, the future horizon is parameterized  by $V$ with $\xi \to 0, t_R\to \infty$, and the past one is parameterized by $U$ with $\xi \to 0, t_R\to -\infty$.
The asymptotic boundary ($\xi \to \infty$) of this wedge is  $\mathbf{R}_{t_R} \times \mathbf{H}^{d-1}$, which can be  mapped by a conformal transformation to the Minkowski-Rindler wedge \cite{Czech:2012be, Casini:2011kv}.\footnote{Extending this conformal map to a coordinate transformation in the bulk, entanglement entropy for the AdS-Rindler wedge for bulk scalar fields is computed in \cite{Sugishita:2016iel}.}

Similarly, the coordinates of the left AdS-Rindler wedge $(t_L,\xi,\chi, \bO)$ are obtained by 
\begin{align}
\begin{split}
    &X^{-1}=\sqrt{1+\xi^2}\cosh \chi, \quad X^0=-\xi \sinh t_L,\quad  
    X^1=-\xi \cosh t_L,\\
    &X^j=\sqrt{1+\xi^2} \sinh \chi\, \hat{y}^j(\bO)\quad (j=2, \dots, d),
\end{split}
\end{align}
where $-\infty<t_L<\infty$. Formally, the left wedge $L$ can be obtained from $R$ by 
$t_L=t_R-i\pi$.

On the right AdS-Rindler wedge, we can express the global coordinates $(\tau,\rho,\theta,\bO)$ in terms of the AdS-Rindler coordinates  $(t_R,\xi,\chi, \bO)$ as
\begin{align}
\label{global-Rindler}
\begin{split}
\tan \tau &=\frac{\xi\sinh t_R}{\sqrt{1+\xi^2}\cosh \chi}, \quad 
    \cos\rho=\frac{1}{\sqrt{(1+\xi^2)\cosh^2 \chi+\xi^2 \sinh^2 t_R}},
    \\
    \tan \theta &=\frac{\sqrt{1+\xi^2}\sinh \chi}{\xi\cosh t_R}.
\end{split}
\end{align}
The asymptotic boundary of the AdS-Rindler patch is a spacetime subregion in the cylinder $\mathbf{R}_\tau \times S^{d-1}$ which is the asymptotic boundary of the global patch as
\begin{align}
\label{bdry:global-Rindler}
\tan \tau =\frac{\sinh t_R}{\cosh \chi}, \quad
    \tan \theta =\frac{\sinh \chi}{\cosh t_R}.
\end{align}
This is a diamond-shaped subregion restricted to $0\leq |\tau\pm\theta|\leq \pi/2$. 
In particular, the asymptotic boundary at $t_R=0$ is a hemisphere $(0\leq \theta\leq \pi/2)$\footnote{For $d=2$, the subregion is in the range $-\pi/2 \leq \theta\leq \pi/2$. } in the time-slice $S^{d-1}$ at $\tau=0$. 
We represent this spacelike subregion in the time-slice $\tau=0$ by $A$, and the spacetime diamond subregion by $D(A)$.
Correspondingly, 
we also represent the bulk time slice  $t_R=0$  in the AdS-Rindler patch by $a$ which is a subregion in the global patch time slice $\tau=0$, and does the region  covered by the AdS-Rindler  patch by $D(a)$.

The diamond subregion  $D(A)$ in the cylinder is conformal to $\mathbf{R}_{t_R} \times \mathbf{H}^{d-1}$ via  \eqref{bdry:global-Rindler} as 
\begin{align}
\label{conf-cyl-Rind}
    -d\tau^2 + d\Omega_{d-1}^2 =e^{2\Phi} \left(
    -dt_R^2+dH^2_{d-1}
    \right),
\end{align}
where we have defined the conformal factor $e^{\Phi}$, which will often appear below, as
\begin{align}
\label{conf-fact}
    e^{\Phi(t_R,\chi)}
    :=\frac{1}{\sqrt{\cosh^2 \chi+\sinh^2 t_R}}=\frac{1}{\sqrt{\cosh u \cosh v}},
\end{align}
where $u=t_R-\chi$, $v=t_R+\chi$.
Note that from \eqref{global-Rindler} we have
\begin{align}
\label{asympt-conf-map}
    \lim_{\xi\to \infty} (\xi \cos \rho)
    =e^{\Phi}.
\end{align}

\subsection{Free scalar fields in AdS-Rindler patch}
Here we summarize the canonical quantization of free scalar field $\phi$ with mass $m$ in the  $(d+1)$-dimensional AdS-Rindler patch.

The equations of motion, $(\Box-m^2) \phi=0$, take the following form in the right AdS-Rindler patch:
\begin{align}
   &\left[- \frac{1}{\xi^2}\partial_{t_R}^2+\frac{1}{\sqrt{-g}}\partial_\xi (\sqrt{-g}(1+\xi^2)\partial_\xi)+\frac{1}{1+\xi^2}\nabla_{H}^2
   -m^2\right]\phi=0
\end{align}
with $\sqrt{-g}=\xi(1+\xi^2)^\frac{d-2}{2}$.
The positive frequency modes are given by
\begin{align}
    v_{\omega,\lambda, \mu}=e^{-i \omega t_R} \tilde{\psi}_{\omega, \lambda}(\xi) Y_{\lambda,\mu}(\chi,\Omega).
\end{align}
Here, $\omega$ is a positive continuous parameter.  $Y_{\lambda,\mu}(\chi,\Omega)$ are the harmonic functions\footnote{We normalize $Y_{\lambda,\mu}(\chi,\Omega)$ such that 
\begin{align*}
    \int_{\mathbf{H}^{d-1}} dV \, Y_{\lambda,\mu}(\chi,\Omega)\, Y^\ast_{\lambda',\mu'}(\chi,\Omega)=\delta(\lambda-\lambda')\delta_{\mu,\mu'}.
\end{align*}} on $\mathbf{H}^{d-1}$, which satisfy \begin{align}
    \nabla_{H}^2 Y_{\lambda,\mu}(\chi,\Omega)
    =-\left[\lambda^2+\left(\frac{d-2}{2}\right)^2\right]Y_{\lambda,\mu}(\chi,\Omega).
\end{align}
$\tilde{\psi}_{\omega, \lambda}(\xi)$ are chosen so that they do not blow up at the boundary $\xi=\infty$, and are given by 
\begin{align}
\label{psiR}
\tilde{\psi}_{\omega, \lambda}(\xi) 
=\frac{N_{\omega,\lambda}}{\Gamma(\nu+1)} \xi^{i\omega} (1+\xi^2)^{-\frac{i\omega}{2}-\frac{\Delta}{2}} 
    ~_2F_1\left(\frac{i\omega-i\lambda+\nu+1}{2} ,\frac{i\omega+i\lambda+\nu+1}{2}
  ;\nu+1;\frac{1}{1+\xi ^2}\right),
\end{align}
where 
\begin{align}
    \Delta:=\frac{d}{2}+\sqrt{m^2+\frac{d^2}{4}}, \qquad 
    \nu:=\Delta-\frac{d}{2}=\sqrt{m^2+\frac{d^2}{4}}.
\end{align}
Note that the right-hand side of \eqref{psiR} is real if we take the normalization constant $N_{\omega,\lambda}$ real (it is invariant under the flip $\omega \to -\omega$).
Near $\xi =0$,  $\tilde{\psi}_{\omega, \lambda}(\xi)$ behave as
\begin{align}
    \tilde{\psi}_{\omega, \lambda}(\xi)\sim N_{\omega,\lambda}\left[
    \frac{\Gamma(-i\omega)\xi^{i\omega}}
    {\Gamma\left(\frac{-i\omega+i\lambda+\nu+1}{2}\right)\Gamma\left(\frac{-i\omega-i\lambda+\nu+1}{2}\right)}
    +
    \frac{\Gamma(i\omega)\xi^{-i\omega}}
    {\Gamma\left(\frac{i\omega-i\lambda+\nu+1}{2}\right)\Gamma\left(\frac{i\omega+i\lambda+\nu+1}{2}\right)
    }
    \right].
    \label{xi_near0}
\end{align}

We fix the real constant $N_{\omega,\lambda}$ in \eqref{psiR}  so that we have
\begin{align}
    (v_{\omega,\lambda, \mu}, v_{\omega',\lambda', \mu'})_R
    =\delta(\omega-\omega')\delta(\lambda-\lambda')\delta_{\mu,\mu'},
\end{align}
where $(\,, \,)_R$ is 
the Klein-Gordon inner product in the AdS-Rindler patch defined as
\begin{align}
    (\phi_1,\phi_2)_R=i\int^{\infty}_0 \!d \xi  \int_{\mathbf{H}^{d-1}} dV  
    \frac{(1+\xi^2)^\frac{d-2}{2}}{\xi}\left(\phi_1^\ast \partial_{t_R} \phi_2-(\partial_{t_R} \phi_1^\ast)\phi_2\right).
\end{align}
This normalization means  
\begin{align}
    \int^{\infty}_0 \!d \xi  
    \frac{(1+\xi^2)^\frac{d-2}{2}}{\xi}
    \tilde{\psi}_{\omega, \lambda}(\xi) \tilde{\psi}_{\omega', \lambda}(\xi)=\frac{1}{2\omega}\delta(\omega-\omega'),
\label{norm_xi0}
\end{align}
and also
\begin{align}
    \int^{\infty}_0 \!d \omega\, 2\omega\,  \tilde{\psi}_{\omega, \lambda}(\xi) 
     \tilde{\psi}_{\omega, \lambda}(\xi') 
     =\frac{\xi}{(1+\xi^2)^\frac{d-2}{2}}\delta(\xi-\xi'). 
     \label{norm_xi}
\end{align}
Evaluating \eqref{norm_xi} at $\xi\sim \xi' \sim 0$ using \eqref{xi_near0}, the normalization constant is fixed as 
\begin{align}
    N_{\omega,\lambda}=\frac{|\Gamma\left(\frac{i\omega-i\lambda+\nu+1}{2}\right)|\, |\Gamma\left(\frac{i\omega+i\lambda+\nu+1}{2}\right)|}{\sqrt{4\pi\omega} |\Gamma(i\omega)|}.
\label{nc}
\end{align}
Note that $\tilde{\psi}_{\omega, \lambda}(\xi)$ in \eqref{xi_near0} behaves as a plain wave with $x= \ln \xi$ near the horizon $x  \sim -\infty $  and the dominant contributions of the integration of $\xi$ in \eqref{norm_xi0} come from 
the region near $x  \sim -\infty $.

For later use, we will evaluate $N_{\omega,\lambda}$ for $\omega \gg 1$ and $|\lambda| \gg 1$.
Using the formula $|\Gamma(i y)| = (\frac{\pi}{y \sinh (\pi y)})^\frac{1}{2}$ 
and $|\Gamma(x+i y)|  \rightarrow \sqrt{2 \pi} y^{x-\frac{1}{2}} e^{-\pi |y|/2}$ for $y \rightarrow \infty$ with $x$ fixed where 
$x,y$ are real,
we find 
$N_{\omega,\lambda} \rightarrow 
 \left( \frac{\omega^2-\lambda^2}{4} \right)^{\frac{\nu}{2}} 
e^{-\frac{\pi}{4} (|\omega-\lambda| +|\omega+\lambda| -2 |\omega| )}$
 in the limit $\omega, |\lambda| \rightarrow \infty$.
Here, we introduce the normalization constant
\begin{align}
N_{\omega,\lambda}^{\rm CFT}
=
\begin{cases}
 \left( \frac{\omega^2-\lambda^2}{4} \right)^{\frac{\nu}{2}}
&\text{for}\quad
\omega^2 \geq \lambda^2\\
0
&\text{for}\quad
\omega^2 < \lambda^2
\end{cases}
\label{ncm}
\end{align}
which naturally appears in large $N$ CFTs on Minkowski space as we will see below. 
Then, we find 
\begin{align}
N_{\omega,\lambda}
\rightarrow
\begin{cases}
N_{\omega,\lambda}^{\rm CFT}
&\text{for}\quad
\omega^2 \geq \lambda^2\\
\left( \frac{\omega^2-\lambda^2}{4} \right)^{\frac{\nu}{2}} e^{-\frac{\pi}{2} (|\lambda|-\omega)}
&\text{for}\quad
\omega^2 < \lambda^2
\end{cases}
\label{N-NCFT}
\end{align}
in the limit $\omega, |\lambda| \rightarrow \infty$.

We can expand the field $\phi$ in the right wedge as
\begin{align}
\label{phi_R}
    \phi(t_R, \xi,\chi, \Omega)=\int d\omega \int d \lambda \sum_\mu \left( a_{\omega,\lambda, \mu} v_{\omega,\lambda, \mu}+a^{\dagger}_{\omega,\lambda, \mu} v^{\ast}_{\omega,\lambda, \mu}\right).
\end{align}
Then, $a_{\omega,\lambda, \mu}$ satisfies
\begin{align}
    [a_{\omega,\lambda, \mu}, a^{\dagger}_{\omega',\lambda', \mu'}]
    =\delta(\omega-\omega')\delta(\lambda-\lambda')\delta_{\mu,\mu'}.
\end{align}

The important point is that the AdS-Rindler energy $\omega$ can take any positive value  independently of $\lambda,\mu$.
Thus, there are modes such that $\omega^2 <\lambda^2$, and we will call them tachyonic modes.\footnote{In AdS$_2$, there are no tachyonic modes \cite{Dey:2021vke}.} 
In the next section, we will argue that the tachyonic modes $(\omega^2 <\lambda^2)$ cannot %be reconstructed from
exist in the CFT on $\mathbf{R}_{t_R} \times \mathbf{H}^{d-1}$ which is the boundary of the AdS-Rindler patch.
We also will see in sec.~\ref{sec:geodesic} that the tachyonic modes mainly constitute the wave packets propagating from the past  horizon to the future one without reaching the asymptotic boundary.

For free theories without UV cutoff, 
the scalar $\phi(t_R, \xi,\chi, \Omega)$ with the AdS-Rindler quantization \eqref{phi_R} is the same as that at the same point with the global quantization.
Then, the reduced density  matrix for the vacuum state of the global-AdS Hamiltonian is the thermal state for the AdS-Rindler Hamiltonian.
However, this is not true if a UV cutoff exists as we argue in the next section. 
In fact, if we consider quantum gravity, the free field description on the fixed background is just an effective theory below a UV cutoff, e.g., the Planck energy.
For holographic CFTs with large but finite $N$ which we are interested in, there must be a UV cutoff in the bulk, and then the transformation between the two quantization (global and AdS-Rindler) is not valid as we will see.

\section{Incompleteness of AdS-Rindler bulk reconstruction}\label{sec:reconstruction}

To simplify the discussion, we focus on $d=2$. 
Then, the asymptotic boundary of the AdS-Rindler patch is (conformal to) $\mathbf{R}^{1,1}$ with metric $ds^2=-dt_R^2+d\chi^2$.  
In this case, the expansion of the bulk scalar in \eqref{phi_R} takes 
\begin{align}
\label{d=2_phi}
    \phi(t_R, \xi,\chi)=\int^{\infty}_{0}d\omega \int^{\infty}_{-\infty} d \lambda \frac{1}{\sqrt{2\pi}}
    \tilde{\psi}_{\omega, \lambda}(\xi) 
    \left[
    a_{\omega,\lambda} e^{-i\omega t_R +i \lambda \chi}+a^{\dagger}_{\omega,\lambda} e^{i\omega t_R -i \lambda \chi}
    \right].
\end{align}
Here, we assumed that this bulk free scalar $\phi(t_R, \xi,\chi)$ is valid even in the UV limit,
i.e. it is UV complete.
We will show below that the tachyonic modes ($\omega^2<\lambda^2$) cannot %be reconstructed from
exist in the  CFT on $\mathbf{R}^{1,1}$. 
To be more precise, the boundary limit of the bulk local operator \eqref{d=2_phi} cannot be the CFT primary field,
and thus the BDHM map fails for the AdS-Rindler case.

The (global) HKLL bulk reconstruction \cite{Hamilton:2006az} is based on the BDHM map \cite{Banks:1998dd}. 
The map relates the asymptotic form of bulk local operator in the global AdS $\phi(\tau, \rho, \theta)$ to a large $N$ CFT operator $O^{CFT}_\Delta$ as
\begin{align}
    \lim_{\rho \to \pi/2} \cos (\rho)^{-\Delta} \phi(\tau, \rho, \theta)=O^{CFT}_\Delta(\tau, \theta),
    \label{BDHM0}
\end{align}
up to a numerical constant. 
For the AdS-Rindler patch, a naive BDHM map would be
\begin{align}
    \lim_{\xi \to \infty} \xi^\Delta \phi(t_R, \xi, \chi)  =O_\Delta(t_R,\chi).
    \label{BDHM}
\end{align}
In fact, it gives the correct conformal transformation of 
%the primary field on the boundary CFT 
the generalized free approximation of $O^{CFT}_\Delta(\tau, \theta)$
for the conformal map \eqref{conf-cyl-Rind}:
$O^{\rm GF}_\Delta(\tau, \theta) = e^{-\Delta \Phi} \, O_\Delta(t_R,\chi)$.
Here $O^{\rm GF}_\Delta(\tau, \theta)$ is the generalized free approximation of 
the primary field on the boundary CFT $O^{CFT}_\Delta(\tau, \theta)$.
This is because $\phi(\tau, \rho, \theta)$ is identified as $\phi(t_R, \xi, \chi)$ in the right AdS-Rindler wedge and 
$\cos\rho \rightarrow \frac{e^{\Phi}}{\xi}$ near the boundary as \eqref{asympt-conf-map}.

In the free bulk theory approximation, using the expansion \eqref{d=2_phi}, $O_\Delta(t_R,\chi)$ can be written as
\begin{align}
    O_\Delta(t_R,\chi)=\int^{\infty}_{0}d\omega \int^{\infty}_{-\infty} d \lambda \frac{N_{\omega,\lambda}}{\sqrt{2\pi}\Gamma(\nu+1)}
    \left[
    a_{\omega,\lambda} e^{-i\omega t_R +i \lambda \chi}+a^{\dagger}_{\omega,\lambda} e^{i\omega t_R -i \lambda \chi}
    \right].
\label{o1}
\end{align}
Then, as done in the original HKLL paper \cite{Hamilton:2006az}, the bulk ladder operators $a_{\omega,\lambda}$ can be expressed by $O_\Delta$ as
\begin{align}
\label{a_by_O}
    a_{\omega,\lambda}=\int^{\infty}_{-\infty}\frac{d t_R}{2\pi} \int^{\infty}_{-\infty} \frac{d \chi}{2\pi} \frac{\sqrt{2\pi}\Gamma(\nu+1)}{N_{\omega,\lambda}}e^{i\omega t_R -i \lambda \chi} O_\Delta(t_R,\chi).
\end{align}
However, $O_\Delta(t_R,\chi)$ has to have modes $e^{-i\omega t_R +i \lambda \chi}$ to obtain nonzero $a_{\omega,\lambda}$.
This is impossible for  $\omega^2<\lambda^2$ if $O_\Delta(t_R,\chi)$ is a CFT primary operator on $\mathbf{R}^{1,1}$ because the existence of such modes implies a tachyonic state, where the mass squared $\omega^2-\lambda^2$ is negative, in the CFT spectrum.
We generally exclude tachyonic states because QFTs containing tachyonic states are problematic. 
For example, by Lorentz transformations, tachyonic states are mapped to states with zero energy but a finite momentum $(\omega=0, \lambda>0)$, and they lead to  an infinite degeneracy of the vacuum state. 
Tachyonic states also contradict with the standard K\"all\'en–Lehmann representation of the two-point function in relativistic QFTs. 
Thus, in standard CFTs on $\mathbf{R}^{1,1}$, local operators do not have modes $e^{-i\omega t_R +i \lambda \chi}$ with $\omega^2<\lambda^2$.
Therefore, CFT on $\mathbf{R}^{1,1}$ cannot reconstruct the ladder operators $a_{\omega,\lambda}$ for the tachyonic modes $(\omega^2 <\lambda^2)$. 

Indeed, because the metric on $D(A)$ is conformal to that on $\mathbf{R}^{1,1}$ by the transformation \eqref{bdry:global-Rindler},
the scalar primary field at a point $(t_R,\chi)$ can be obtained by the conformal transformation of the primary field $O^{CFT}_\Delta(\tau, \theta)$ on the cylinder as \begin{align}
    O^\text{CFT,flat}_{\Delta}(t_R,\chi)
    :=e^{\Delta \Phi}O^{CFT}_\Delta(\tau, \theta).
\end{align}
We know the large $N$ spectrum of the holographic CFT on the Minkowski space $\mathbf{R}^{1,1}$. 
The primary field $O^\text{CFT,flat}_{\Delta}(t_R,\chi)$ for the large $N$ CFT on $\mathbf{R}^{1,1}$ should be the same as that obtained by the HKLL reconstruction on the Poincare patch:
\begin{align}
    O^\text{CFT,flat}_{\Delta}(t_R,\chi)
    =\int^{\infty}_{|\lambda|}d\omega \int^{\infty}_{-\infty} d \lambda \frac{N^{\rm CFT}_{\omega,\lambda}}{\sqrt{2\pi}\Gamma(\nu+1)}
    \left[
    a^{\rm CFT}_{\omega,\lambda} e^{-i\omega t_R +i \lambda \chi}+{a^{\rm CFT}_{\omega,\lambda}}^{\dagger} e^{i\omega t_R -i \lambda \chi}
    \right],
\label{o2}
\end{align}
where $a^{\rm CFT}_{\omega,\lambda}$ are also normalized annihilation operators and $N^{\rm CFT}_{\omega,\lambda}$ is defined in \eqref{ncm}. 
What is important here is that $O^\text{CFT,flat}_{\Delta}$ in \eqref{o2}
does not contain tachyonic modes.
Thus, we conclude $O_\Delta(t_R,\chi) \neq O^\text{CFT,flat}_{\Delta}(t_R,\chi)$ and 
the Hilbert spaces for these two operators are completely different.

Note that %by construction, 
the two point function of the global AdS, $ \langle 0|  T ( O^{CFT}_\Delta (\tau, \theta) \, O^{CFT}_\Delta(\tau', \theta') ) | 0 \rangle$, can be reproduced by 
%the $O_\Delta (t_R,\chi)$
$O^\text{CFT,flat}_{\Delta}(t_R,\chi)$ 
on the corresponding points as
\begin{align}
      \langle 0|  T ( O^{CFT}_\Delta (\tau, \theta) \, O^{CFT}_\Delta(\tau', \theta') ) | 0 \rangle 
=   e^{-\Delta( \Phi(t_R,\chi)+\Phi(t'_R,\chi'))}
\tr_A ( \rho_{A} T ( O^\text{CFT,flat}_{\Delta} (t_R,\chi) \,  O^\text{CFT,flat}_{\Delta}(t'_R, \chi') ) ),
\label{np1}
\end{align}
where $\ket{0}$ is the CFT vacuum on the cylinder, and $\rho_{A}=\tr_{\bar{A}} (| 0 \rangle \langle 0| )$ is the reduced density matrix in the CFT Hilbert space on $A$.
This is just the usual relation between 
the CFT on the cylinder and that on the Rindler subregion.\footnote{
If we consider the Poincare $AdS_3$, instead of the global $AdS_3$, 
the map between the two coordinates for the CFT is just the two dimensional Minkowski-Rindler map and 
the corresponding $\rho_{A}$ is known to be the thermal density matrix. 
}
The two point function can be also approximately reproduced by $O_\Delta (t_R,\chi)$ 
in the low energy region
as
\begin{align}
      \langle 0|  T ( O^{CFT}_\Delta (\tau, \theta) \, O^{CFT}_\Delta(\tau', \theta') ) | 0 \rangle 
      \simeq & e^{-\Delta( \Phi(t_R,\chi)+ \Phi(t'_R,\chi'))}
\tr_a ( \rho_{a}  T ( O_\Delta(t_R,\chi)  \, O_\Delta(t'_R,\chi')  ) ),
\label{np2}
\end{align}
where $\rho_{a}=\tr_{\bar{a}} (| 0, \text{bulk} \rangle \langle 0, \text{bulk} |  )$ is the density matrix in the bulk free scalar Hilbert space on $a$, with the vacuum $| 0, \text{bulk} \rangle$ in the global AdS.
This comes from the usual relation between 
the bulk scalar on the global AdS and the one on the AdS-Rindler subregion,
$ \langle 0|  T (\phi  (\tau, \rho, \theta) \,\phi (\tau', \rho',\theta') ) | 0 \rangle
=\tr_a ( \rho_{a}  T ( \phi (t_R,\xi, \chi)  \, \phi (t'_R,\xi',\chi')  ) )$,
with the boundary limit of the points of the operator insertions.\footnote{
We assumed the expected completeness of 
the modes of the (UV complete) bulk free field in the left and right Rindler wedges.
}
Then, the $n$-point functions are also reproduced in the large $N$ limit by the factorization.
We also note that the AdS-Rindler HKLL reconstruction \cite{Hamilton:2006az} 
with treating the smearing function as a distribution \cite{Morrison:2014jha} works
well. 
However, the ``CFT'' operators used there are constructed from the bulk local operators by 
the naive BDHM map \eqref{BDHM}, i.e. $O_\Delta(t_R,\chi)$, which 
is different from the CFT operators
$O^\text{CFT,flat}_{\Delta}(t_R,\chi)$.
It is worth emphasizing that the correlation functions of $O_\Delta(t_R,\chi) $ and $O^\text{CFT,flat}_{\Delta}(t_R,\chi)$
are different although \eqref{np2} holds.
Indeed, it is obvious that 
\begin{align}
\langle 0_a |  T ( O_\Delta(t_R,\chi) \, O_\Delta(t'_R,\chi') ) | 0_a \rangle
\neq
\langle 0_A |  T ( O^\text{CFT,flat}_{\Delta}(t_R,\chi) \, O^\text{CFT,flat}_{\Delta} (t'_R,\chi') ) | 0_A  \rangle,
\end{align}
where $ | 0_a \rangle$ is the vacuum of the bulk theory in the region $a$ and $ | 0_A \rangle$ is the vacuum of the CFT in the region $A$ (i.e. the Minkowski vacuum on $\mathbf{R}^{1,1}$),
because the coefficient $N_{\omega,\lambda}$ in \eqref{o1} is different from $N^{\rm CFT}_{\omega,\lambda}$ in \eqref{o2}
and the Hilbert spaces are different even in the low energy.
The equation \eqref{np2} is valid only for the special states $\rho_a, \rho_A$ as the low-energy approximation.
In particular, the bulk correlation function $\langle 0_a |  T ( \phi(t_R,\xi,\chi) \, \phi(t'_R,\xi',\chi') ) | 0_a \rangle$ cannot be reproduced from $O^\text{CFT,flat}_{\Delta}(t_R,\chi)$ because of the lack of tachyonic $a_{\omega,\lambda}$
with $\omega< |\lambda|$.

\paragraph{What is wrong with $O_\Delta(t_R,\chi) $?}

Because $O_\Delta(t_R,\chi)$ is obtained by the conformal transformation from $O^{\rm GF}_\Delta(\tau,\theta)$ which is the generalized free approximation of $O^{CFT}_\Delta(\tau,\theta)$,
it seems natural to assume that
$O_\Delta(t_R,\chi) = O^\text{CFT,flat}_{\Delta}(t_R,\chi)$ in the low energy, and indeed
it has been assumed, in particular to consider the subregion duality, the entanglement wedge reconstruction and the error correction code in the holographic theory.
The reason why it is violated is that 
the generalized free theory is the large $N$ limit approximation and 
such a spectrum is only the low energy approximation and not realized for the high energy states.
This is clear for states with the Planck energy which correspond to black holes.
This means that the CFT operator $ O^{CFT}_\Delta(\tau, \theta)$ obtained from the
bulk local operators in the global AdS by the BDHM map is 
not correct for the high energy modes.
In particular, the (high momentum) tachyonic modes in $O_\Delta(t_R,\chi)$ are composed by such nonexistent high energy modes
of the global AdS or the CFT on the cylinder, and then they are absent in the CFT in $D(A)$.

In general, we claim that
the low energy states and operators in the Rindler patch depend on the UV completion of the theory,
which implies that the quantum gravity effects are important for them if the theory includes the gravity.
In order to see this, let us first consider the free scalar field with mass $m$ in 
$d+1$-dimensional Minkowski spacetime $ds^2=-dt^2+dx^2 +dy^i dy_i$ where $i=1,2,\cdots, d-1$, instead of 
the scalar fields in AdS because  these two models are very similar for the aspects discussed here. 
(A difference is that the ``tachyonic'' modes are not special for the Minkowski case.)
The usual right Rindler patch is given by $t_R=\tanh^{-1} (t/x),\, \zeta=\ln \sqrt{x^2-t^2}$.
We denote the ladder operators associated with the global modes $e^{i(-\sqrt{k^2+k_i k^i+m^2} t+ k x+ k_i y^i)}$  as  $a_{k, k_i}$ where $k$ is the momentum in $x$-direction and those with the right Rindler modes $e^{i(-\omega t_R + k_i y^i)} K_{i \omega} (\sqrt{k_i k^i+m^2}\zeta)$ as $a_{\omega,  k_i}$.
The Bogoliubov transformation $a_{\omega, k_i}= 
\sum_{k,k'_i} (\alpha^*_{\omega,  k_i ;k,k'_i} a_{k,k'_i}+\beta^*_{\omega, k_i ;,k,k'_i} a^\dagger_{k,k'_i})$
is known (see e.g. \cite{Crispino:2007eb}) to be
\begin{align}
\alpha_{\omega,  k_i ;k,k'_i} = \frac{\prod_i \delta(k_i-k'_i)}{\sqrt{2 \pi \sqrt{k^2+k_i k^i+m^2} (1-e^{- 2 \pi \omega})}}
\left( \frac{\sqrt{k^2+k_i k^i+m^2}-k}{\sqrt{k_i k^i+m^2}}\right)^{i \omega},
\end{align}
and $\beta_{\omega,  k_i ;k,k'_i} = e^{-\pi \omega} \alpha_{\omega,  k_i ;k,k'_i} $.
For large $|k|$, we can approximate 
\begin{align}
\alpha_{\omega,  k_i ;k,k'_i} \sim \frac{\prod_i \delta(k_i-k'_i)}{ \sqrt{2 \pi (1-e^{- 2 \pi \omega})}} \frac{1}{|k|^\frac{1}{2}}
e^{i \omega (-\ln(|k|) +\ln(\sqrt{k_i k^i+m^2}/2))}
\end{align}
for $k>0$ and
$\alpha_{\omega,  k_i ;k,k'_i} \sim \frac{\prod_i \delta(k_i-k'_i)}{ \sqrt{2 \pi (1-e^{- 2 \pi \omega})}} \frac{1}{|k|^\frac{1}{2}}
e^{i \omega (\ln(|k|) -\ln(\sqrt{k_i k^i+m^2}/2))}$ for $k<0$.
Let us consider the ``norm'' of $\alpha_{\omega,  k_i ;k,k'_i}$ in the global vacuum $| 0 \rangle$:
\begin{align}
\frac{\tr_R ( e^{- 2 \pi H_R}  a_{\omega, k_i} a_{\omega, k_i}^\dagger) }{\tr_R ( e^{- 2 \pi H_R} )}
=
\langle 0| (\sum_{k,k'_i} \alpha^*_{\omega,  k_i ;k,k'_i} a_{k,k'_i}) (\sum_{\tilde{k},\tilde{k}'_i} \alpha^*_{\omega,  k_i ;\tilde{k},\tilde{k}'_i} a_{\tilde{k},\tilde{k}'_i})^\dagger| 0 \rangle
=\sum_{k,k'_i} |\alpha_{\omega,  k_i ;k,k'_i}|^2,
\end{align}
where $H_R$ and $\tr_R$ are the Hamiltonian and  the trace of the right Rindler wedge.
More precisely, we smear $k_i$ directions  (and $\omega$ direction later) of $a_{\omega, k_i}$, 
for example by the Gaussian factor, as $ \int dk_i e^{-\frac{1}{2 \epsilon^2} (k^i-\bar{k}^i) (k_i-\bar{k}_i)} a_{\omega, k_i}$. Then, coefficients become non-singular as
$\alpha_{\omega;k,k'_i} \equiv \int dk_i e^{-\frac{1}{2 \epsilon^2} (k^i-\bar{k}^i) (k_i-\bar{k}_i)} \alpha_{\omega,  k_i ;k,k'_i} 
\sim \frac{e^{-\frac{1}{2 \epsilon^2}  ({k'}^i-\bar{k}^i) (k'_i-\bar{k}_i) }}{ \sqrt{2 \pi (1-e^{- 2 \pi \omega})}} \frac{1}{|k|^{\frac{1}{2}}}
e^{i \omega (\ln(|k|) -\ln(\sqrt{k'_i {k'}^i+m^2}/2))}$.
We can see that, for large $|k|$, $\sum_{k} |\alpha_{\omega;k,k'_i}|^2$ behaves as 
$\sum_{k} \frac{1}{|k|}$ which is divergent.\footnote{
This divergence is regularized by the smearing: $ \int dk_i e^{-\frac{1}{2 \epsilon^2} (k^i-\bar{k}^i) (k_i-\bar{k}_i)-\frac{R^2}{2} (\omega-\bar{\omega})^2} a_{\omega, k_i}$.
Then,
the ``norm'' behaves like $\sum_{k} \frac{1}{|k|} e^{-\frac{ 1}{2 R^2} (\ln |k|)^2} \sim R$
where the smearing of the energy is very small $1/R \ll 1$ where $R$ may be regarded as an IR regularization.
(Here, we neglect the $1/ (1-e^{- 2 \pi \omega})$ factor by taking a large $\bar{\omega}$.)
The dominant contributions are from $\ln |k| \sim R$.
Note that if we take $1/R \gg 1$, the contributions from the high momentum and energy modes are 
negligible. For the localized wave packets, we need to take $1/R \gg 1$.
This implies that we can neglect such modes for the (smeared) local operators apart from the horizon.
}
This implies that the Rindler mode $a_{\omega, k_i}$ cannot be constructed if we neglect the 
global modes with arbitrary high momentum and energy.
Thus, if the free scalar theory is a low energy effective theory,
(even the low energy sector of) the spectrum of the theory on the Rindler wedge depends on the UV completion of the theory.
In particular, if the theory couples to a gravity, it depends on quantum effects of the gravity,
which are specified by the dual CFT for the AdS/CFT case.

The property that the low energy Rindler modes contain arbitrary high energy modes is reminiscent of the brick wall proposal  \cite{tHooft:1984kcu, Iizuka:2013kma}, where a divergence in the large $N$ limit arises from the near-horizon behaviors of fields.
We expect that this is the essential nature of (black hole) horizons. In order to see this, first let us define the lightcone coordinates:
$\tilde{u}=t-x, \tilde{v}=t+x$ and $u=t_R-\zeta, v=t_R+\zeta$,
in which the Rindler horizon are $|u| \rightarrow \infty$ or $|v| \rightarrow \infty$.
The relations between these are
\begin{align}
\tilde{u}=-e^{- u}, \,\, \tilde{v}=e^{v},
\end{align}
which implies that 
$\delta \tilde{u}=e^{- u} \delta u, \,\, \delta \tilde{v}=e^{v} \delta v$
where $\delta u$, for example, means small variation of $u$.
Thus, near the Rindler horizon (for example, $u \gg 1$)
a lightcone momentum ($\sim 1/\delta u$) in the Rindler patch corresponds to a large lightcone momentum ($\sim 1/\delta \tilde{u}$) in the global coordinates with the ratio $e^{u} \gg 1$.
This explains why 
the low energy Rindler modes contain arbitrary high energy modes intuitively.
Furthermore, this is expected to be a universal property of horizons
and actually a similar problem happens for the AdS-Rindler horizon as we will see.

Let us return to the AdS/CFT case
and consider the Bogoliubov transformation between the global AdS modes 
$a^{\rm global}_{n m}$
and 
the AdS-Rindler modes $a_{\omega,\lambda}$. 
We will see that AdS-Rindler modes contain 
infinitely high momentum global AdS modes as the above Minkowski case. 
Intuitively, the problem comes from behaviors of fields near the AdS-Rindler horizon as similar to the Minkowski-Rindler case.

The bulk local operator $\phi(\tau, \rho, \theta) $
can be expanded by the modes in the global AdS $a^{\rm global}_{n m}$ as
\begin{align}
\phi(\tau, \rho, \theta) =
\sum_{n,m} 
\left( 
a^{ \rm global \,\, \dagger}_{nm} \, e^{i \omega_{n m} \tau}  e^{-i m \theta}
+a^{\rm global}_{n m} e^{-i \omega_{n m} \tau}  e^{i m \theta}
\right)
\psi_{nm}^\text{bulk}(\rho)
\end{align}
where $n$ is a non-negative integer, $m$ is an integer,
\begin{align}
\omega_{nm}=2n+|m|+\Delta,
\end{align}
and $\psi_{nm}^\text{bulk}(\rho)$ are modes in $\rho$-direction whose explicit form is not used here.\footnote{It is given by $\psi_{nm}^\text{bulk}(\rho)=\frac{1}{{\cal N}_{nm}} \sin^{|m|} (\rho) \cos^\Delta (\rho)\,
P_n^{|m|, \Delta-1} \left( \cos(2 \rho) \right)$, where  
${\cal N}_{nm}$ is the numerical constant given in \cite{Fitzpatrick:2010zm} and $P_n^{|m|, \Delta-1}$ is the Jacobi polynomial.}
Here, we will use the following relation between the 
$O_\Delta(t_R,\chi) $ and $O^{CFT}_{\Delta}(\tau,\theta)$
which should give the correct Bogoliubov coefficients:
\begin{align}
    a_{\omega,\lambda}
&=\frac{ \Gamma(\nu+1)}{\sqrt{2\pi} N_{\omega,\lambda}}
\int^{\infty}_{-\infty }dt_R \int^{\infty}_{-\infty} d \chi 
 e^{i\omega t_R -i \lambda \chi} O_\Delta(t_R,\chi) \\
&=\frac{ \Gamma(\nu+1)}{\sqrt{2\pi} N_{\omega,\lambda}}
\int^{\infty}_{-\infty }dt_R \int^{\infty}_{-\infty} d \chi 
 e^{i\omega t_R -i \lambda \chi} 
e^{\Delta \Phi(t_R,\chi)} \, O^{CFT}_\Delta(\tau,\theta),
\label{bogo}
\end{align}
where $ O^{CFT}_\Delta(\tau,\theta)$ can be expanded \cite{Terashima:2020uqu, Terashima:2021klf} as 
\begin{align}
 O^{CFT}_\Delta(\tau,\theta) =
\sum_{n,m} 
\psi_{n m}^{CFT} \, 
\left( 
a^{ \rm global \,\, \dagger}_{nm} \, e^{i \omega_{n m} \tau}  e^{-i m \theta}
+a^{\rm global}_{n m} e^{-i \omega_{n m} \tau}  e^{i m \theta}
\right),
\end{align}
with
\begin{align}
\psi_{n m}^{CFT} \, 
=
\sqrt{\frac{2 \Gamma(n+\Delta) \Gamma(n+|m|+\Delta)}
{\pi \Gamma(\Delta)^2  \Gamma(n+1) \Gamma(n+|m|+1)}}.
\end{align}
Then, the coefficient of $a^{\rm global}_{n m}$ in the expansion of $a_{\omega,\lambda}$,
\eqref{bogo}, is 
\begin{align}
 \frac{ \Gamma(\nu+1) \psi_{n m}^{CFT}}{\sqrt{2\pi} N_{\omega,\lambda}}
\int^{\infty}_{-\infty }dt_R \int^{\infty}_{-\infty} d \chi 
 e^{i\omega t_R -i \lambda \chi  -i \omega_{n m} \tau +i m \theta } 
(\cosh^2 \chi + \sinh^2 t_R)^{-\Delta/2},
\label{coef1}
\end{align}
where $\tau=\tan^{-1 }\frac{\sinh t_R}{\cosh \chi}, \,\, \theta=\tan^{-1} \frac{\sinh \chi}{\cosh t_R}$.
Let us concentrate on high momentum and energy modes, i.e. $\omega \gg1, |\lambda| \gg1$.
Then, the integrals in \eqref{coef1} almost vanish because of the phase cancellation
except the region in which the phase is almost a constant.
Using the coordinates $u =t_R - \chi, v= t_R+ \chi$ instead of $t_R,\chi$, the region of stationary phase is given by the conditions\footnote{
The derivative of the other factor in \eqref{coef1} is 
$\delta \ln((\cosh u \cosh v)^{-\Delta/2})=-(\Delta/2)(\tanh u\,  \delta u+\tanh v\, \delta v) $
in which $\tanh u, \tanh v$ are not large.
}
\begin{align}
\begin{split}
0= \partial_u(\omega t_R-\lambda \chi  - \omega_{n m} \tau + m \theta)
   =\frac{1}{2}\left( \omega +\lambda - \frac{\omega_{nm} +m}{\cosh u}\right)\,,\\
   0= \partial_v(\omega t_R-\lambda \chi  - \omega_{n m} \tau + m \theta)
   =\frac{1}{2}\left( \omega -\lambda - \frac{\omega_{nm} -m}{\cosh v}\right)\,.
\end{split}
\label{coef2}
\end{align}
% The conditions \eqref{coef2} can be written\footnote{
% The derivative of the other factor in \eqref{coef1} is 
% $\delta \ln((\cosh u \cosh v)^{-\Delta/2})=-(\Delta/2)(\tanh u\,  \delta u+\tanh v\, \delta v) $
% in which $\tanh u, \tanh v$ are not large.
% }
% \begin{align}
% \begin{split}
%      0 = \omega -\lambda - \frac{\omega_{nm} -m}{\cosh u},\\
% 0=  \omega +\lambda - \frac{\omega_{nm} +m}{\cosh v}.
% \end{split}
% \label{coef3}
% \end{align}
%In particular, we can evaluate $\gamma_\pm $ for the regions near the Rindler horizon. 
%For $|t_R| \gg 1, |t_R| \gg |\chi|$
%we find $\gamma_\pm \sim 2 e^{-|t_R| \pm {\rm sgn} (t_R) \chi}$ and 
%for $ |\chi| \gg 1,  |\chi| \gg |t_R|$,
%we find $\gamma_\pm \sim 2 e^{-|\chi| \pm {\rm sgn} (\chi) t_R}$.
%For $|u| \gg 1, |u| \gg |v|$ 
%we find $\gamma_+ \sim  1, \,\, \gamma_- \sim  e^{-|u|} \cosh v$, and,
%for $|v| \gg 1, |v| \gg |u|$ 
%we find $\gamma_- \sim  1, \,\, \gamma_+ \sim  e^{-|v|} \cosh u$.

% For $\omega^2-\lambda^2 \ge 0$,
% the equations \eqref{coef3}
% are (approximately) solved for any (large) $n,m$ if $\omega_{nm} \pm m \ge \omega \pm \lambda$
% at appropriate points $\{ t_R,\chi \}$.
% In particular, for $n^2 \gg \omega^2+\lambda^2, m^2 \gg \omega^2+\lambda^2$
% the corresponding points are $|u| \gg 1$ and $|v| \gg 1$ 
% except the case that $2n-|m|$ is small.
For $\omega^2-\lambda^2 \ge 0$,
the conditions \eqref{coef2} almost always have solutions at appropriate points  $\{ u,v\}$ for any (large) $n,m$. 
In particular, for $n^2 \gg \omega^2+\lambda^2, m^2 \gg \omega^2+\lambda^2$
the corresponding points are $|u| \gg 1$ and $|v| \gg 1$  except the case that $n-|m|$ is small.
For the generic case, the stationary points are given by
$e^{|u|} \sim  \frac{\omega_{nm} +m}{\omega +\lambda}$ and $e^{|v|} \sim  \frac{\omega_{nm} -m}{\omega -\lambda}$.
For the exceptional case, the corresponding points are $|v| \gg 1$ as
$e^{|v|} \sim  \frac{\omega_{nm} -m}{\omega -\lambda}$ with finite $u$
for $m<0$ or $|u| \gg 1$ as $e^{|u|} \sim  \frac{\omega_{nm} +m}{\omega +\lambda}$ with finite $v$ for $m>0$.
(These points %which correspond to very large $\omega_{nm}, |m|$ 
are near the Rindler horizon.)

Let us evaluate \eqref{coef1} using the stationary phase approximation.
We can see from \eqref{coef2} that the second derivatives of the phase 
${\omega t_R - \lambda \chi  - \omega_{n m} \tau + m \theta }$
is
$\frac{1}{2} ((\omega_{nm} +m ) \frac{\sinh u}{\cosh^2 u} {\delta u}^2+(\omega_{nm} -m ) \frac{\sinh v}{\cosh^2 v}{\delta v}^2) $ which is approximated to 
$\frac{1}{2} ((\omega +\lambda) \tanh u ( \delta u)^2+(\omega -\lambda ) \tanh v ( \delta v)^2) $ near the stationary points.
%$\frac12 ((\omega +\lambda) ( \delta v)^2+(\omega -\lambda ) ( \delta u)^2) $ for $|u| , |v| \gg 1$,
%$\frac12 ((\omega +\lambda) \tanh v ( \delta v)^2+(\omega -\lambda )  ( \delta u)^2) $ %for $|u| \gg 1$ and
%$\frac12 ((\omega +\lambda) ( \delta v)^2 +(\omega -\lambda ) \tanh u  ( \delta u)^2) $ for $|v| \gg 1$.
They do not depend on $\omega_{nm}, m$ in the limit $|m| \rightarrow \infty$.
We can also see that
$\psi_{n m}^{CFT} \sim \sqrt{n(n+|m|)}^{\Delta-1}$ for $n, |m| \gg 1$ and 
$\psi_{n m}^{CFT} \sim \sqrt{|m|}^{\Delta-1}$ for $|m| \gg 1$.
Finally,  the factor $(\cosh u \cosh v)^{-\Delta/2}$ behaves like $e^{-\frac{\Delta}{2}(|u|+|v|)}$ for $|u| , |v| \gg 1$,
$e^{-\frac{\Delta}{2} |u|}$ for $|u| \gg 1$,
or $e^{-\frac{\Delta}{2} |v|}$ for $|v| \gg 1$.
Therefore, the $(m,n)$-dependence of the coefficient \eqref{coef1} is evaluated as $1/{(n(n+|m|))}^{1/2}$  for large $n,|m|$ 
and $1/{|m|}^{1/2}$  for large $|m|$. 
Note that both of $\sum_{n,m} 1/{(n(n+|m|))}$ and $\sum_{m} 1/|m|$  are divergent.

It is also important to note that the phase factor which depends on both $|m|$ and $\omega,\lambda$ is approximately 
 $i\omega t_R -i \lambda \chi  = i (   (\omega +\lambda) u+(\omega -\lambda ) v  )/2  \sim i (   (\omega +\lambda) \ln (\omega_{nm} +m)+(\omega -\lambda )  \ln(\omega_{nm} -m) )/2   $
for $|u|, |v| \gg 1$.
This is also similar to the Minkowski case.
Thus, the mode $a^{\rm global}_{n m}$ with an arbitrary large $\omega_{nm}, |m|$
contributes to $a_{\omega,\lambda}$ as we have seen for the Minkowski-Rindler case.

For $\omega^2-\lambda^2 < 0$ (tachyonic), the condition in \eqref{coef2},
 $0 = \omega -\lambda - (\omega_{nm} -m)/\cosh v$, cannot be 
solved even approximately because $\omega_{nm} -m \ge 0$  where we 
take $\lambda >0$ for simplicity.
Thus, coefficients \eqref{coef1} are exponentially suppressed by $|\lambda|-\omega$ for any $n,m$ 
which may be consistent with the fact that $N_{\omega,\lambda}$ is exponentially small as \eqref{N-NCFT}.
In this case, the largest contributions for $\lambda>0$ are from 
\begin{align}
\begin{split}
0&= \omega +\lambda - (\omega_{nm} +m)/\cosh u, \\
 0 &=  - (\omega_{nm} -m) /\cosh v .
\end{split}
\end{align}
Thus, the mode $a^{\rm global}_{n m}$ with an arbitrary large $\omega_{nm}, |m|$
contributes to $a_{\omega,\lambda}$ because these are same as the above 
by formally substituting $\omega-\lambda=0$.
Note that for this tachyonic mode, 
these large $\omega_{nm}, |m|$ modes are the dominant contributions.
For these tachyonic modes, as we have seen, there are no saddle points for the integration for the overlap between the global and Rindler modes. Thus, the high energy global modes contribute dominantly. This indicates that the tachyonic  modes are composed by these modes which do not exist in the low energy effective theory. This matches with the fact that the tachyonic modes do not exist in the CFT.

\paragraph{Reconstructable operators}

The bulk local field $\phi  (\tau, \rho, \theta)$ is equivalent to $\phi (t_R,\xi, \chi) $
in the right AdS-Rindler wedge if we assume $\phi  (\tau, \rho, \theta)$ is valid for the UV limit.
More precisely, as the bulk free theory without a UV cutoff, we have the formal operator equality, $\phi  (\tau, \rho, \theta)=\phi (t_R,\xi, \chi)  \otimes 1_{\bar{a}}$ in the right AdS-Rindler patch, where $1_{\bar{a}}$ is the identity operator on the left AdS-Rindler patch.
We can extract $\phi (t_R,\xi, \chi)$ by a projection as $\phi (t_R,\xi, \chi)={\rm P}_a (\phi  (\tau, \rho, \theta))$ where $P_a(\mathcal{O})= \tr_{\bar{a}} ( {\cal O} )/ \tr_{\bar{a}} (1_{\bar{a}})$.\footnote{We can instead $\phi (t_R,\xi, \chi) = \langle 0_{\bar{a}} |  
\phi  (\tau, \rho, \theta) | 0_{\bar{a}}  \rangle $
because $\phi  (\tau, \rho, \theta)$ is linear in the creation and annihilation operators.} 

However, as explained, the bulk operator $\phi (t_R,\xi, \chi)= {\rm P}_a (\phi  (\tau, \rho, \theta))$ cannot be reconstructed from CFT on $A$.
The reconstructable part of the bulk local field in the right Rindler wedge might be
\begin{align}
 \phi^\text{part} (t_R,\xi, \chi) = {\rm P}_A (\phi  (\tau, \rho, \theta)),
\end{align}
where 
$ {\rm P}_A({\cal O} )= \tr_{\bar{A}} ( {\cal O} )/ \tr_{\bar{A}} (1)$,
which also  remove the identity operator in the Hilbert space on $\bar{A}$ like ${\rm P}_a$ in above.
This satisfies the following BDHM like relation,
\begin{align}
    \lim_{\xi \to \infty} \xi^\Delta \phi^\text{part} (t_R,\xi, \chi)={\rm P}_A  ( e^{\Delta\Phi} O^{CFT}_\Delta(\tau, \theta) )
=O^\text{CFT,flat}_{\Delta}(t_R,\chi),
    \label{BDHM3}
\end{align}
because $O^{CFT}_\Delta(\tau, \theta) $ is a CFT operator supported on the region $A$ on which ${\rm P}_A$ acts trivially.
Here we assume that the mode expansion of ${\rm P}_A (\phi (t_R,\xi, \chi) )$ with the modes ${\rm P}_A (a_{\omega,\lambda}) $ can be used 
%after the projection ${\rm P}_A$
although it is not fully justified.
Then, we find the following BDHM like relation also,
\begin{align}
    \lim_{\xi \to \infty} \xi^\Delta \phi^\text{part} (t_R,\xi, \chi)
    ={\rm P}_A ( \lim_{\xi \to \infty} \xi^\Delta \phi (t_R,\xi, \chi) )
    ={\rm P}_A ( O_\Delta(t_R,\chi) ).
    \label{BDHM2}
\end{align}
Equating these two relations, 
we find 
$ {\rm P}_A ( a_{\omega,\lambda} )
=\frac{ N^{\rm CFT}_{\omega,\lambda} }{N_{\omega,\lambda}}  a^{\rm CFT}_{\omega,\lambda} $
by noticing that the both of $a^{\rm CFT}_{\omega,\lambda}$ and $ a_{\omega,\lambda}$ have 
the same energy $\omega$ and momentum $\lambda$ conjugate to $t_R, \chi$.\footnote{Note also that ${\rm P}_A ( a_{\omega,\lambda})$ approximately identified with $a^{\rm CFT}_{\omega,\lambda}$ for non-tachyonic high momentum and energy modes because of \eqref{N-NCFT}.}
%If we throw away the horizon-horizon modes in \eqref{d=2_phi} as 
Then, we find
\begin{align}
\label{d=2_phi_part}
    \phi^\text{part}(t_R, \xi,z)=\int^{\infty}_{-\infty} d \lambda \int^{\infty}_{|\lambda|}d\omega  \frac{1}{\sqrt{2\pi}}
    \frac{ N^{\rm CFT}_{\omega,\lambda} }{N_{\omega,\lambda}} \tilde{\psi}_{\omega, \lambda}(\xi) 
    \left[
    a^{\rm CFT}_{\omega,\lambda} e^{-i\omega t_R +i \lambda \chi}+{a^{\rm CFT}_{\omega,\lambda}}^{\dagger} e^{i\omega t_R -i \lambda \chi}
    \right],
\end{align}
where the tachyonic modes are absent.
This bulk operator $\phi^\text{part}$ can be reconstructed from the CFT.
%because \eqref{a_by_O} works for the modes $\omega^2>\lambda^2$.  
The expression is
\begin{align}
\label{d=2_phi_part2}
    \phi^\text{part}(t_R, \xi,z)=&\int^{\infty}_{-\infty} d \lambda \int^{\infty}_{|\lambda|}d\omega 
    \int^{\infty}_{-\infty}\frac{d t'_R}{2\pi} \int^{\infty}_{-\infty} \frac{d \chi'}{2\pi} 
    f_{\omega,\lambda}(\xi)
    \nn
    &\times
    \left[e^{-i\omega (t_R-t'_R) +i \lambda (\chi-\chi')}+e^{i\omega (t_R-t'_R) -i \lambda (\chi-\chi')}\right]O^\text{CFT,flat}_\Delta(t'_R,\chi')
\end{align}
where 
\begin{align}
    f_{\omega,\lambda}(\xi)=\xi^{i\omega} (1+\xi^2)^{-\frac{i\omega}{2}-\frac{\Delta}{2}} 
    ~_2F_1\left(\frac{i\omega-i\lambda+\nu+1}{2} ,\frac{i\omega+i\lambda+\nu+1}{2}
  ;\nu+1;\frac{1}{1+\xi ^2}\right).
\end{align}

We then define the smearing function as
\begin{align}
\label{smearingK}
    K(t'_R,\chi'; t_R,\xi,\chi):=\int^{\infty}_{-\infty} d \lambda \int^{\infty}_{|\lambda|}d\omega  \left[e^{-i\omega (t_R-t'_R) +i \lambda (\chi-\chi')}+e^{i\omega (t_R-t'_R) -i \lambda (\chi-\chi')}\right]
    f_{\omega,\lambda}(\xi).
\end{align}
This smearing function makes sense unlike the smearing function in \cite{Hamilton:2006az} where the integral contains the region $\lambda^2 \gg \omega^2$ and diverges. 
To overcome the divergence, an analytic continuation to complex $\chi$ is proposed in \cite{Hamilton:2006az}. 
It is also proposed in \cite{Morrison:2014jha} to treat the smearing function as a distribution.
We do not need these treatments because we do not have the problematic region $\lambda^2 \gg \omega^2$ in our integral \eqref{smearingK}.
Thus, we can interchange the order of integration over $(\lambda, \omega)$ and $(t'_R,\chi')$ in  \eqref{d=2_phi_part2}, 
and obtain
\begin{align}
\label{d=2_reconst}
    \phi^\text{part}(t_R, \xi,z)=
    \int^{\infty}_{-\infty}\frac{d t'_R}{2\pi} \int^{\infty}_{-\infty} \frac{d \chi'}{2\pi} K(t'_R,\chi'; t_R,\xi,\chi) O^\text{CFT,flat}_\Delta(t'_R,\chi').
\end{align}

It is important to note that
$\phi^\text{part}(t_R, \xi,z)$ is not a bulk local operator because it does not contain a part of bulk modes. 
Hence, we cannot reconstruct bulk local operators from the CFT on $\mathbf{R}^{1,1}$.
The missing modes are tachyonic modes. 
As we will see in Sec.~\ref{sec:geodesic}, these modes give the main contribution of the wave packets propagating from the past horizon to the future one without reaching the AdS boundary. 
It means that the CFT on the subregion $D(A)$ in the cylinder cannot reconstruct these wave packets. 
Of course, if we use CFT operators on the entire cylinder, we can construct bulk operators, and thus can describe such wave packets from the CFT.

The incompleteness also means that the CFT reduced density matrix on the subregion $A$ of the entire time-slice of the cylinder is not dual to the bulk reduced density matrix on the time-slice $a$ of the AdS-Rindler patch, $\rho_A^\text{CFT} \neq \rho_a^\text{bulk}$, even in the low energy region.
In order to obtain the correspondence, 
we need to regard operators generated by \eqref{d=2_reconst} as the bulk operators. 
It is not precise, but 
if we trace-out the tachyonic modes in the bulk theory on the Rindler wedge,
the density matrices becomes similar
as
\begin{align}
    \rho_A^\text{CFT} \sim \tr_{\omega^2<\lambda^2} \rho_a^\text{bulk},
\end{align}
because $\omega, |\lambda| \gg 1$ the modes can be approximately identified.
In this sense, a weak version \cite{Terashima:2020uqu, Terashima:2021klf} of the subregion duality holds: any low-energy operators supported only on the boundary subregion $D(A)$ can be described by low-energy bulk ones on $D(a)$, although the inverse is not possible. 

The identification rule is schematically summarized as follows. 
First, the low energy Hilbert space $\mathcal{H}^\text{CFT}$ of the large $N$ CFT on the cylinder can be identified with the bulk one $\mathcal{H}^\text{bulk}$ on the global AdS as $\mathcal{H}^\text{CFT} = \mathcal{H}^\text{bulk}$.
Those Hilbert spaces can be formally decomposed into the tensor products of the subregions as $\mathcal{H}^\text{CFT} = \mathcal{H}^\text{CFT}_A \otimes \mathcal{H}^\text{CFT}_{\bar{A}}$ and $\mathcal{H}^\text{bulk} = \mathcal{H}^\text{bulk}_a \otimes \mathcal{H}^\text{bulk}_{\bar{a}}$.
Then, our claim is $\mathcal{H}^\text{CFT}_A  \neq \mathcal{H}^\text{bulk}_a$. 
Nevertheless, $\mathcal{H}^\text{CFT}_A  \subset \mathcal{H}^\text{bulk}_a$ holds, and we may obtain a subregion identification $\mathcal{H}^\text{CFT}_A  = P_A( \mathcal{H}^\text{bulk}_a)$  by reducing the bulk space $\mathcal{H}^\text{bulk}_a$ by the projection $P_A$ or tracing-out the tachyonic modes.

\paragraph{Higher dimensions}
We can also show that the tachyonic modes ($\omega^2<\lambda^2$) are absent in CFT on higher-dimensional space $\mathbf{R}_{t_R} \times \mathbf{H}^{d-1}$.
It is clear for high frequency modes $\omega, \lambda \gg 1$ because we can ignore the curvature of $\mathbf{H}^{d-1}$ for these modes, and modes $\omega^2- \lambda^2<0$ are tachyonic in $\mathbf{R}^{1,d-1}$.

For general frequencies, we use the fact that the spacetime
$\mathbf{R}_{t_R} \times \mathbf{H}^{d-1}$ is conformally flat.
CFTs on the conformally flat spacetime can be described by the CFTs on Minkowski spacetime
because of the traceless property of the energy momentum tensor which couples to the Weyl factor of the metric variation.
For the free CFT case, we can see it explicitly.
The d'Alembert operator $\Box_{\mathbf{R^{1,d-1}}}$ in the Minkowski spacetime is mapped, up to the conformal factor,  to
\begin{align}
\label{eq:box}
   \Box_{\mathbf{R}_{t_R} \times \mathbf{H}^{d-1}}- \xi_\text{conf} R,
\end{align}
where the last term is the conformally curvature coupling term with 
\begin{align}
    \xi_\text{conf}=\frac{d-2}{4(d-1)},
\end{align}
and $R$ is the Ricci scalar for $\mathbf{R}_{t_R} \times \mathbf{H}^{d-1}$ given by $R=-(d-1)(d-2)$.
The modes  $e^{-i \omega t_R} Y_{\lambda,\mu}(\chi,\Omega)$ are eigenmodes of the operator \eqref{eq:box} as
\begin{align}
    \left[\Box_{\mathbf{R}_{t_R} \times \mathbf{H}^{d-1}}- \xi_\text{conf} R \right]e^{-i \omega t_R} Y_{\lambda,\mu}(\chi,\Omega)
    =(\omega^2-\lambda^2)e^{-i \omega t_R} Y_{\lambda,\mu}(\chi,\Omega).
\end{align}
Thus, modes with $\omega^2-\lambda^2<0$ are tachyonic, and should be absent in CFT on $\mathbf{R}_{t_R} \times \mathbf{H}^{d-1}$.
Therefore, the holographic CFT cannot reconstruct the tachyonic modes.

\paragraph{Global AdS to Poincare AdS}
Instead of the Rindler patch of the global AdS, 
we can consider the Poincare patch of the global AdS
and consider corresponding CFT.
This CFT corresponding to the Poincare patch should be CFT on Minkowski space because the metric on the boundary is conformally Minkowski space.
For the Poincare patch, it is well known that there are no tachyonic modes 
in the bulk free theory in the Poincare patch. 
The BDHM map on the patch gives the CFT operator
$O^{CFT}_\Delta(t,x)$ of \eqref{o2} in the Minkowski space.
Thus, the situation is different from the Rindler patch.
We can also see that 
the bulk modes in the Poincare patch 
do not depend on very high momentum modes
in the global AdS, as we will see below.

The Poincare patch for $AdS_3$ in the global $AdS_3$ is defined by
\begin{align}
    t=\frac{\sin \tau}{\cos \tau + \cos \theta \sin \rho}, \,\,\,
    x=\frac{\sin \theta \sin \rho}{\cos \tau + \cos \theta \sin \rho}, \,\,\,
    z=\frac{\cos \rho}{\cos \tau + \cos \theta \sin \rho}, 
\end{align}
where $-\infty < t, x, < \infty$ and $0 <z < \infty$.
On the boundary $z \rightarrow 0$,
they become
\begin{align}
    t=\frac{\sin \tau}{\cos \tau + \cos \theta }, \,\,\,
    x=\frac{\sin \theta }{\cos \tau + \cos \theta }.
\end{align}
Using the lightcone coordinates  $u =t -x, v= t+x$ and
 $\tilde{u} =\tau-\theta, \tilde{v}= \tau+\theta$,
the relations can be written as
\begin{align}
   u=\tan \frac{\tilde{u}}{2}, \,\,\,
v=\tan \frac{\tilde{v}}{2},
\end{align}
which implies that $d \tilde{u} = \frac{2}{1+u^2} du$ and $d \tilde{v} = \frac{2}{1+v^2} dv$.
The metric on the boundary is given by $ds^2=d \tilde{u}  d \tilde{v}= \frac{4}{(1+u^2)(1+v^2)} du dv $.
We denote the annihilation operator associated with the mode $e^{i\omega t -i \lambda x}$ in the Poincare patch by $a^P_{\omega,\lambda}$.
Then, 
as 
\eqref{coef1} for the Rindler case, 
the coefficient of $a^{\rm global}_{n m}$ in the expansion of $a^P_{\omega,\lambda}$
is 
\begin{align}
 \frac{2^{\Delta} \Gamma(\nu+1) \psi_{n m}^{CFT}}{\sqrt{2\pi} N^{\rm CFT}_{\omega,\lambda}}
\int^{\infty}_{-\infty }dt \int^{\infty}_{-\infty} d x\,
 e^{i\omega t -i \lambda x  -i \omega_{n m} \tau +i m \theta } 
((1+u^2)(1+v^2))^{-\Delta/2}.
\label{coef1a}
\end{align}
The phase cancellation occurs at
\begin{align}
0= \omega +\lambda - (\omega_{nm} +m)  \frac{2}{1+u^2}, \,\,\,\,
 0 = \omega -\lambda - (\omega_{nm} -m) \frac{2}{1+v^2}.
\label{coef3a}
\end{align}
Thus, qualitatively the Poincare case is very similar to the Rindler case. 
However, the important differences are
that there are no tachyonic modes $\omega^2 < \lambda^2$ in the Poincare patch
and 
the phase factor which depends on both $|m|$ and $\omega,\lambda$ is approximately 
 $i\omega t -i \lambda x  = i (   (\omega +\lambda) u+(\omega -\lambda ) v )/2  \sim i (   \sqrt{ (\omega +\lambda)  (\omega_{nm} +m)}+\sqrt{ (\omega -\lambda )  (\omega_{nm} -m)} )/(2\sqrt{2})$ for $|u|, |v|\gg1$.
This implies that the smearing of $\omega$ by an IR regularization $R$
gives a cut-off of the high momentum modes as  $|m| < R, \, \, n < R$.

\section{Null geodesics in the AdS-Rindler patch}\label{sec:geodesic}

In this section, we see that in the AdS-Rindler patch there are null geodesics never reaching the asymptotic boundary. 
This type of null geodesics starts from the past AdS-Rindler horizon and ends on the future one. 
The existence of null geodesics  never reaching the asymptotic boundary is a characteristic difference from the global AdS where all null geodesics reaching the asymptotic boundary. 
We will show that the non-reconstructable modes  $\omega^2<\lambda^2$ in \eqref{phi_R} are related to the horizon-horizon geodesics, and also  modes  $\omega^2>\lambda^2$ are to geodesics reaching the asymptotic boundary. 
Thus, the existence of the horizon-horizon geodesics is a reason why the AdS-Rindler reconstruction is incomplete unlike the global case.
The relation between the null geodesics and the bulk reconstruction is also discussed in \cite{Bousso:2012mh}.

Let us find the null geodesics in the AdS-Rindler patch, 
where the coordinates are $(t_R,\xi, \chi, \bO)$ as summarized in subsec.~\ref{subsec:cdnt}.
For simplicity, we consider only geodesics with constant $\bO$.
Then, when we solve the geodesic equation, we can effectively regard that the geodesic moves in a three-dimensional space as $(t_R(s),\xi(s), \chi(s))$ where $s$ is an affine parameter of the geodesic, and also can regard that $\partial_\chi$ is a Killing vector. 
Thus, along the geodesic, we have the following two conserved quantities $\omega, \lambda$ corresponding to two Killing vectors $\partial_{t_R}, \partial_\chi$ as
\begin{align}
\label{eq:consts}
\tilde{\omega}=\xi^2 \frac{d t_R}{ds}, \quad \tilde{\lambda}=(1+\xi^2) \frac{d \chi}{ds}.
\end{align}
We set the ratio of the two quantities as 
\begin{align}
    b:=\frac{\tilde{\lambda}}{\tilde{\omega}}.
\end{align}
The geodesic also has to satisfy the null condition 
\begin{align}
    -\xi^2\left(\frac{d t_R}{ds}\right)^2+\frac{\left(\frac{d \xi}{ds}\right)^2}{1+\xi^2}
    +(1+\xi^2)\left(\frac{d \chi}{ds}\right)^2
    =0.
\end{align}
Combining these equations, we obtain
\begin{align}
\label{eq:geods}
    \left(\frac{d \xi}{dt_R}\right)^2=\xi^2 \left[1+\left(1-b^2\right)\xi^2\right].
\end{align}
The behavior of the geodesic depends on whether the ratio $|b|$ is greater than 1 or not. 

\paragraph{Case (i): $|b|>1$.}
In this case, the range of $\xi$ must be in $0\leq \xi \leq \frac{1}{\sqrt{b^2-1}}$ so that the right hand side of \eqref{eq:geods} is positive. 
Thus, any geodesics with $|b|>1$ can never reach the asymptotic boundary $\xi=\infty$.
Eq.~\eqref{eq:geods} is easily solved as
\begin{align}
    \xi(t_R)=\frac{1}{\sqrt{b^2-1}\cosh(t_R-t_0)},
\end{align}
where $t_0$ is an integration constant.
This geodesic comes from the past horizon and  goes to the future horizon without reaching the boundary. 
We also show that 
$\chi(t_R)$ is given by
\begin{align}
    \chi(t_R)=\chi_0+\frac{1}{2}\log \frac{b+\tanh(t_R-t_0)}{b-\tanh(t_R-t_0)},
\end{align}
where $\chi_0$ is an integration constant.

\paragraph{Case (ii): $|b|=1$.}
In this case, the geodesics are given by
\begin{align}
    \xi(t_R)=e^{\pm(t_R-t_0)}, \quad \chi(t_R) =\chi_0+\frac{t_R-t_0}{2}\pm \frac{\log \cosh (t_R-t_0)}{2}.
\end{align}
For the upper sign, the geodesic starts from the past horizon and ends on the asymptotic boundary. Similarly, for the lower sign, it starts from the boundary and ends on the future horizon.

\paragraph{Case (iii): $|b|<1$.}
In this case, the solution of \eqref{eq:geods} is 
\begin{align}
    \xi(t_R)=\frac{1}{\sqrt{1-b^2}|\sinh(t_R-t_0)|},
\end{align}
Thus, it comes from the horizon at $t_R=-\infty$ and reaches the boundary at a finite time, 
or 
it starts from the boundary at a time and approaches the horizon at $t_R=\infty$. We also note that $\chi(t_R)$ is given by
\begin{align}
    \chi(t_R)=\chi_0+\frac{1}{2}\log \frac{1+b\tanh(t_R-t_0)}{1-b\tanh(t_R-t_0)}.
\end{align}

Therefore, in the AdS-Rindler patch, there are null geodesics never reaching the asymptotic boundary [Case (i)].
We will call these the horizon-horizon geodesics, while the other geodesics [Case (ii) and (iii)] the boundary-horizon geodesics.

Let us consider scalar waves well localized on null geodesics and see what modes dominantly contributes to the waves. 
The mass of scalar, $m$, is negligible because we use the geometrical optics approximation supposing that the mode frequencies $\omega, \lambda$ are sufficiently larger than the mass and the curvature scale of AdS, $\omega, \lambda \gg m, \ell_\text{AdS}^{-1}=1$. 

As an example, we consider $d=2$ below. 
Near a point $(t_R,\xi,\chi)=(t_0, \xi_0,\chi_0)=x^\mu_0$ on a null geodesics, we introduce the locally flat coordinates as
\begin{align}
    \bar{t}=\xi_0 t_R, \quad \bar{\xi}=\frac{\xi}{\sqrt{1+\xi_0^2}}, \quad 
    \bar{\chi}=\sqrt{1+\xi_0^2}\chi. 
\end{align}
In the coordinates, the velocity of the $\bar{\chi}$-direction at the point $x^\mu_0$ is
\begin{align}
\label{eq:chi-v}
   \bar{v}_\chi= \frac{d\bar{\chi}}{d\bar{t}}=\frac{\sqrt{1+\xi_0^2}}{\xi_0}\frac{d\chi}{dt}=\frac{b\xi_0}{\sqrt{1+\xi_0^2}},
\end{align}
where we have used \eqref{eq:consts}.
The velocity of the $\bar{\xi}$-direction is $\bar{v}_\xi=\pm \sqrt{1+(1-b^2)\xi_0^2}$. 

Let us determine the dominant mode for wave-packet localized to this geodesic. 
It is enough to use the locally flat coordinates because we supposed that the mode frequencies are sufficiently large so that geometrical optics approximation is valid.
We can use plain waves in the locally flat coordinates for the mode expansion at least near $x^\mu_0$. 
The dominant modes propagating along the geodesic should have the velocities $\bar{v}_\xi, \bar{v}_\chi$ and are given by
\begin{align}
    e^{-i\bar{\omega} (\bar{t}-\bar{v}_\xi \bar{\xi}-\bar{v}_\chi \bar{\chi})}
\end{align}
near $x^\mu_0$.
In particular, it has the component
\begin{align}
\label{dom-mode}
     e^{-i\bar{\omega} (\bar{t}-\bar{v}_\chi \bar{\chi})}
     =e^{-i\bar{\omega}\xi_0( t_R-b \chi)}.
\end{align}
Comparing it to the mode expansion \eqref{d=2_phi} with label $\omega, \lambda$, 
we conclude that
the wave well-localized to the null geodesic with the parameter $b$ consists of modes satisfying $\lambda/\omega=b$.
Thus, the sign $\omega^2-\lambda^2$ corresponds to that of $1-|b|$.
The tachyonic modes $\omega^2<\lambda^2$ are the  dominant contributions to the horizon-horizon geodesics.
Therefore, the CFT on the asymptotic boundary of the AdS-Rindler patch cannot describe the bulk wave packets localized to these geodesics.

\section{View from the global AdS}

We have seen that the wave packets localized to the boundary-horizon geodesics consist mainly of modes $\omega^2>\lambda^2$, and those localized to the horizon-horizon geodesics do mainly of modes $\omega^2<\lambda^2$. 
Combining the discussion in section \ref{sec:reconstruction},
we conclude that the CFT on the subregion $D(A)$ can describe the bulk wave packets localized to boundary-horizon geodesics but cannot do those localized to horizon-horizon geodesics.

Instead of using the AdS-Rindler coordinate,
we can study which part of the bulk local operators are able to be reconstructed by CFT operators in 
the subregion $D(A)$ from the global AdS (and the corresponding CFT on the cylinder) viewpoint.
Indeed, in \cite{Terashima:2020uqu, Terashima:2021klf}
such studies have been done 
and the above conclusion was already obtained.
Below, we will shortly review this view from the global AdS.

First, it is easy to see that the bulk local operator cannot be reconstructed from 
the CFT operators supported on  a subregion $D(A)$, where $A$ is a subspace of the whole  space $S^{d-1}$, except $A=S^{d-1}$.
This implies that the subregion duality is not correct as follows.
Let us consider the bulk local operator at $\rho=0$, which is the center of the AdS
on $t=0$ slice
and a boundary subregion $D(A)$ whose causal wedge contains this center.
Then, the bulk local operator at the center are rotational symmetric, 
where we can smear the operator keeping the rotational symmetry.\footnote{
The low energy subspace of the states is invariant under the rotation
because
the rotation commutes with the Hamiltonian.
Thus, the bulk local operator, which is a low energy operator, can be taken rotation invariant. 
}
This rotational symmetry is identified as the rotational symmetry of the CFT,
and the CFT operator which reconstructs the bulk operator should be rotational symmetric.
It means that the operator is homogeneous in the whole space $S^{d-1}$.
On the other hand, it is obvious that the CFT operators supported on a subregion $D(A)$
cannot be rotational symmetric and  homogeneous.
Thus, in order to reconstruct the bulk local operator at the center, we need 
the CFT operators on whole $S^{d-1}$.
We note here that the center is not a special point of AdS because an arbitrary point can be moved to the center by an AdS isometry.
Thus, the reconstruction of any bulk local operators requires 
the CFT operators on whole $S^{d-1}$.\footnote{
If we take the boundary limit of the bulk local operator, it becomes the CFT primary operator at the point.}

Next, we will consider which part of the bulk local operator can be reconstructed from a boundary subregion.
In order to do,
it is useful to consider which bulk well-localized wave packets can be reconstructed
from CFT operators supported on a subregion $D(A)$.
In \cite{Terashima:2020uqu, Terashima:2021klf} using the 
bulk reconstruction developed in 
\cite{Terashima:2017gmc,Terashima:2019wed, Nagano:2021tbu},
a CFT description of bulk well-localized wave packets was given.
For the holographic CFT on the cylinder $\mathbf{R}_\tau \times S^{d-1}$, let us 
consider states well localized to a very small subregion $B$ in $S^{d-1}$ at a time $\tau=\tau_0$. 
For example, $B$ is a ball region around $\Omega=\Omega_0$ as $|\Omega-\Omega_0|<\epsilon_B$. 
Such states are written (in the Schr\"{o}dinger picture) as
\begin{align}
\label{phiB;tau0}
    \ket{\phi_B, \tau_0}=\int d^{d-1}\Omega\, f_B(\Omega) \mathcal{O}_{\Delta}(\Omega)\ket{0}
\end{align}
where $\ket{0}$ is the vacuum state for the Hamiltonian $H$ associated with time $\tau$, and $f_B(\Omega)$ is a smearing function so that only the subregion $B$ is relevant for the integration in \eqref{phiB;tau0}, e.g., the Gaussian distribution centered at a point $\Omega=\Omega_0 \in B$.\footnote{
\label{foot:smear}
To avoid the divergence coming from the local state, we also need the smearing for the time direction, such that the length scale of it should be much smaller than $\epsilon_B$.
We also assume that the support of the smearing function (or the width of the Gaussian) is larger than the length scale of the UV cutoff, which will be the Planck scale, so that high energy states do not appear in  \eqref{phiB;tau0}. 
} 
The time-evolved state of $\ket{\phi_B, \tau_0}$ is given by
\begin{align}
    \ket{\phi_B, \tau}=\int d^{d-1}\Omega\, f_B(\Omega)e^{-iH(\tau-\tau_0)} \mathcal{O}_{\Delta}(\Omega)\ket{0} .
\end{align}
It was shown in \cite{Terashima:2020uqu, Terashima:2021klf} that the above CFT state $\ket{\phi_B, \tau}$ for a very small subregion $B$\footnote{$B$ is small but should be larger than the length scale of the UV cutoff as noted in footnote~\ref{foot:smear}.}
represents a state well localized to a wave packet  moving in the radial direction ($\rho$-direction) in the bulk.
In particular, supposing $-\pi/2<\tau_0<0$ and the center of $B$ is at $\Omega=\Omega_0$, the bulk wave packet almost-localized to a radial-directed null geodesic starting from the boundary point $\Omega=\Omega_0$ at $\tau=\tau_0$ and passing the point $(\rho=\pi/2+\tau_0,\Omega=\Omega_0)$ at $\tau=0$.\footnote{
Here, what we really construct is a sum of the bulk wave packets moving in the radial direction with all energies below the UV cut-off and the averaging over $\Omega_0$ in $S^{d-1}$ approximately gives the bulk local state at $\rho=\tau=0$ for $\tau_0=-\pi/2$, i.e.
$ \int d^{d-1}\Omega\, e^{-iH(\tau+\pi/2)} \mathcal{O}_{\Delta}(\Omega)\ket{0} \sim \phi(\rho=0) \ket{0}$ up to a numerical constant.
Thus, we can regard \eqref{phiB;tau0} as 
a component of a decomposition of the bulk local state.}
Similarly, we can also reconstruct bulk wave packets along other-directed null geodesics starting from a boundary (almost) local region (see \cite{Terashima:2020uqu, Terashima:2021klf} for details).

What is important here is that 
bulk (almost) local states along a bulk null geodesics can be represented 
by 
(almost) local states on a small region $B$ in CFT, if the geodesics 
starting from the boundary region $B$. 
This leads to that, 
we can reconstruct the bulk wave packets along null geodesics 
from CFT states which have the support on a subregion $C$ at $\tau=0$
if the null geodesics reach $D(C)$, which is the domain of dependence of $C$.
%What is important here is that (almost) local states on a small region $B$ in CFT represents bulk (almost) local states along the bulk null geodesics starting from the boundary region $B$. 
%This leads to that, 
%from CFT states which have the support on a subregion $C$ at $\tau=0$, we cannot reconstruct the bulk wave packets along null geodesics such that the geodesics do not reach $D(C)$, which is the domain of dependence of $C$. 
Conversely, we cannot reconstruct the bulk wave packets from the CFT states on $C$
if the null geodesics do not reach $D(C)$
because the state which is given at the small region $B$ is not within $C$ at $\tau=0$.
This is consistent with the fact that we cannot reconstruct the horizon-horizon geodesics in the AdS-Rindler patch $D(a)$ from the CFT on the associated subregion $D(A)$
because the horizon-horizon geodesics do not reach $D(A)$.

%\section{Discussions}

\section*{Acknowledgement}

We thank T. Takayanagi for helpful discussions.
SS acknowledges support from JSPS KAKENHI Grant Number JP 21K13927.
This work was supported by JSPS KAKENHI Grant Number 17K05414.
This work was supported by MEXT-JSPS Grant-in-Aid for Transformative Research Areas (A) ``Extreme Universe'', No. 21H05184.

%%%%%%%%%%%%%%%%%%%%%%%%%%%%%%%%%%%%%%%%%%%%%%%%%%%%%%%%%%%%%%
%%%%%%%%%%%%%%%%%%%%%%%%%%%%%%%%%%%%%%%%%%%%%%%%%%%%%%%%%%%%%%
\bibliographystyle{utphys}
\bibliography{ref-AdSCFT}
%%%%%%%%%%%%%%%%%%%%%%%%%%%%%%%%%%%%%%%%%%%%%%%%%%%%%%%%%%%%%%
\end{document}